\documentclass[twocolumn]{revtex4}
\usepackage[dvips]{graphicx}
\begin{document}

\title{
Thermal conductivity of the thermoelectric layered cobalt oxides
measured by the Harman method}
\author{ 
A. Satake, H. Tanaka, and T. Ohkawa
}

\affiliation{
Department of Applied Physics, Waseda University,
3-4-1 Ohkubo, Shinjuku-ku, Tokyo, 169-8555, Japan
}

\author{\def\thefootnote{\alph{footnote})}
T. Fujii and I Terasaki\footnote{
Author to whom correspondence should be addressed; electronic mail: terra@waseda.jp}
}

\affiliation{
Department of Applied Physics, Waseda University,
3-4-1 Ohkubo, Shinjuku-ku, Tokyo, 169-8555, Japan and
CREST, Japan Science and Technology Agency, Tokyo 103-0027, Japan
}
\begin{abstract}
In-plane thermal conductivity of the thermoelectric layered cobalt oxides
has been measured using the Harman method, in which thermal conductivity 
is obtained from temperature gradient induced by applied current.
We have found that the charge reservoir block (the block other than the 
CoO$_2$ block) dominates the thermal conduction,
where a nano-block integration concept is effective for material design.
We have further found that the thermal conductivity shows a small but finite
in-plane anisotropy between $a$ and $b$ axes, which can be ascribed to 
the misfit structure.
\end{abstract}

\maketitle

Thermoelectric materials have recently attracted a renewed interest 
as a energy-conversion material
in harmony with our environments. 
They are characterized by the dimensionless figure of merit 
$ZT=S^2T/\rho\kappa$,
where $S$, $\rho$ and $\kappa$ are the thermopower, resistivity
and thermal conductivity at temperature $T$, respectively.
In other words, the thermoelectric material is a material
that shows large thermopower, low resistivity and low thermal
conductivity simultanously.

\begin{figure}[b]
\begin{center}
 \includegraphics[width=8cm,clip]{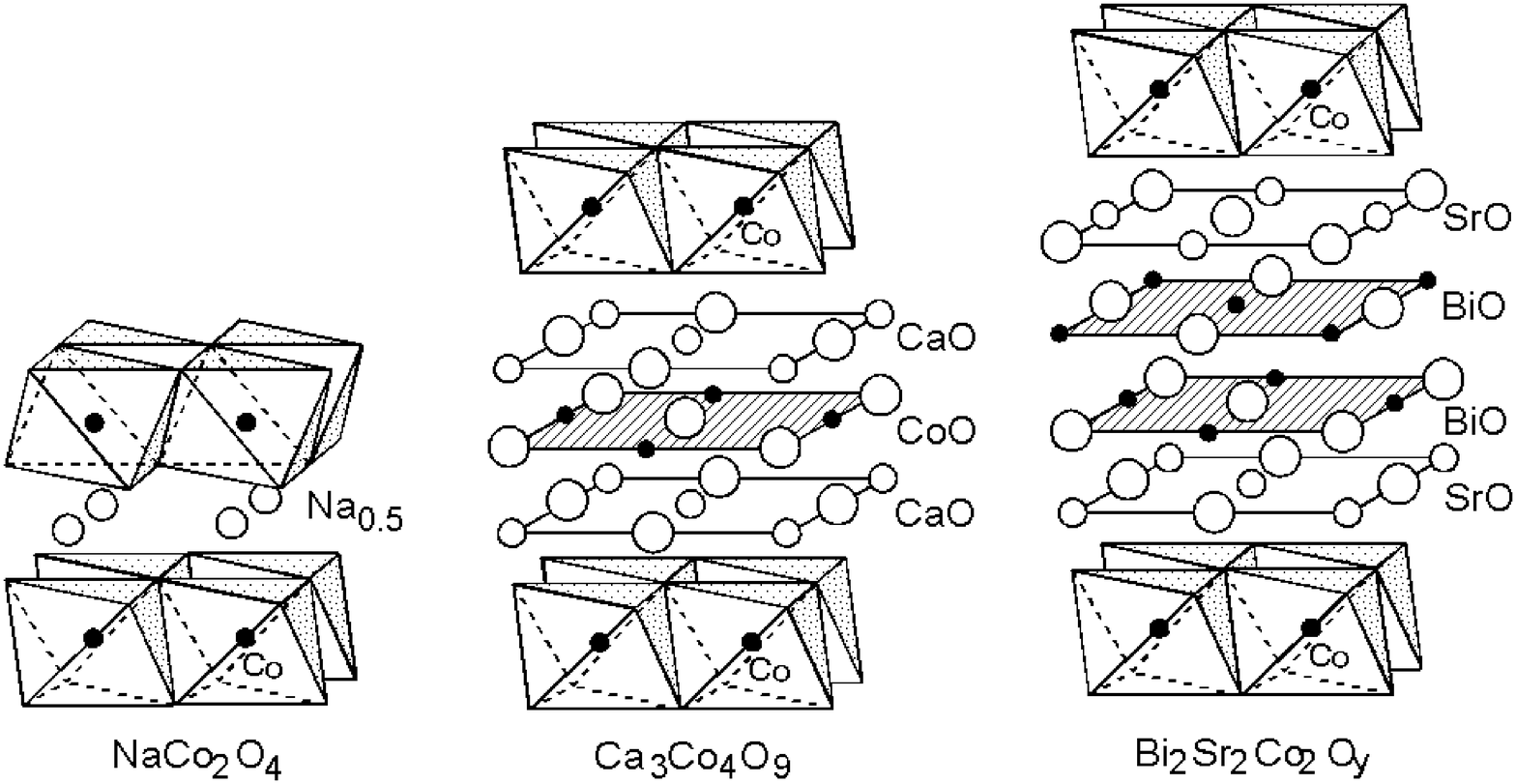}\\ 
\end{center}
\caption{
Crystal structure of the layered cobalt oxides.
}
\label{fig1}
\end{figure}

Oxides were regarded as unsuitable for thermoelectrics,
but the layered cobalt oxides have been extensively investigated
as a promising thermoelectric material
since the discovery of  a large thermopower
in NaCo$_2$O$_4$ (Na$_{0.5}$CoO$_2$) \cite{terra,terra2}.
Soon after, the related layered cobalt oxides
Bi$_{2-x}$Pb$_x$Sr$_2$Co$_2$O$_y$ \cite{itoh,funahashiBiCo}, 
Ca$_3$Co$_4$O$_9$ \cite{miyazaki,masset,co225}, 
Ca$_2$(Cu,Co)$_2$Co$_4$O$_9$ \cite{miyazaki2}
TlSr$_2$Co$_2$O$_y$ \cite{hebert}, 
and (Pb,Co)Sr$_2$Co$_2$O$_y$ \cite{maignan} 
have been synthesized and identified.
They share the hexagonal CdI$_2$-type CoO$_2$ block as a common unit
as shown in Fig. 1, 
and show good thermoelectricity of $ZT>1$ above 1000 K \cite{fujita,kappaBiCo,kappaCa}. 
This means that their thermoelectric performance 
is comparable with that of the conventional thermoelectric materials 
such as PbTe and Si$_{1-x}$Ge$_x$.

Single crystals of the thermoelectric layered cobalt oxides are
very thin and small, and the heat loss is too serious to measure 
$\kappa$ precisely.
There are only a few reports on $\kappa$ measurement of single crystals,
all of which were done above room temperature \cite{fujita,kappaBiCo,kappaCa}.
Fujita et al. \cite{fujita} found that the in-plane thermal conductivity
$\kappa$ of a Na$_x$CoO$_2$ single crystal 
was 200 mW/cmK at room temperature, and decreased rapidly at high temperatures.
This is highly inconsistent with 
the other layered cobalt oxides \cite{kappaBiCo,kappaCa}
and the polycrystalline NaCo$_2$O$_4$ \cite{takahata,takahata2},
which needs cross-checking by other technique.
We have succeeded in measuring in-plane $\kappa$ of single crystals of
Na$_x$CoO$_2$, Bi$_{2-x}$Pb$_x$Sr$_2$Co$_2$O$_y$ and Ca$_3$Co$_4$O$_9$
by the Harman method.
Here we will report on the experimental details of the Harman method and
the in-plane $\kappa$ of the layered cobalt oxides below room temperature.

Bi$_{2-x}$Pb$_x$Sr$_2$Co$_2$O$_y$ single crystals were grown 
by a traveling-solvent floating-zone method \cite{fujii}.
Single crystals of Na$_x$CoO$_2$ and Ca$_3$Co$_4$O$_9$ were grown by the 
flux technique \cite{terra,kappaCa}.
A typical dimension of the samples was 2.0$\times$0.5$\times$0.05 mm$^3$.
Detailed growth conditions and charge transport were given 
in the cited references.

\begin{figure}[t]
\begin{center}
 \includegraphics[width=8cm,clip]{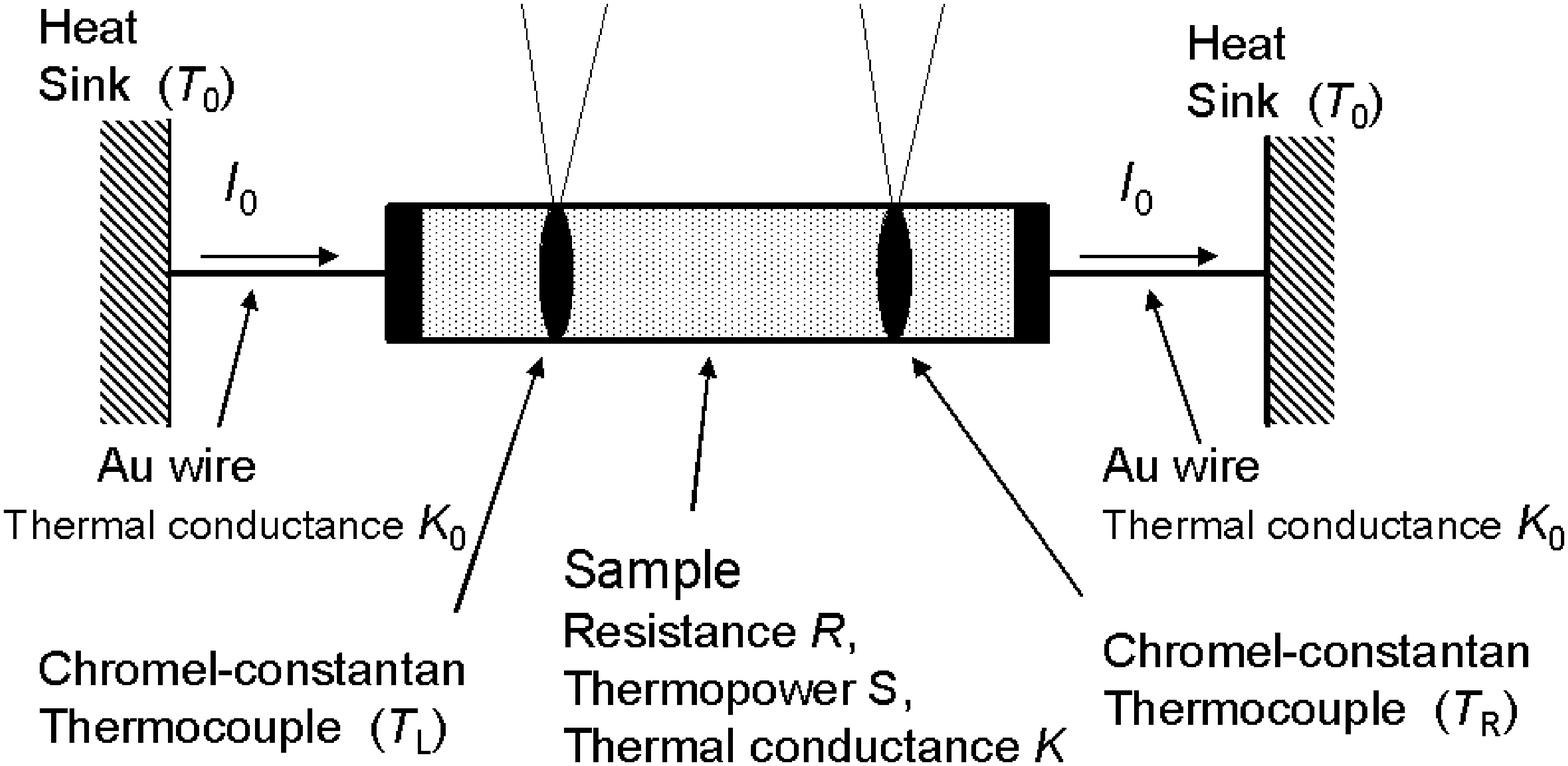}\\ 
\end{center}
\caption{
The sample configuration of the Harman method.
}
\label{fig2}
\end{figure}

Thermal conductivity was measured 
in a closed-cycle refrigerator from 15 to 300 K
by the Harman method, 
where $\kappa$ was obtained from temperature difference 
induced by applied current $I_0$.
Figure 2 shows the experimental configuration.
This is similar to a four-probe method, but
instead of the voltage terminals,
two sets of chromel-constantan thermocouple are pasted to detect the 
temperatures $T_L$ and $T_R$, and the difference $\Delta T=T_R-T_L$.
The sample is hung with current leads (gold wire of 20$\mu$m-diameter)
in vacuum, and is nearly isolated thermally.
Then heat absorption/radiation due to the Peltier effect 
induced by $I_0$ balances a heat backflow  across the sample, and
in a steady state, 
\begin{equation}
STI_0=K \Delta T
\end{equation}
is realized.
Thus the thermal conductance $K$ is obtained 
from $S$ (measured in advance in a different run), $T$ and $\Delta T$.
In a steady-state method, $\kappa$ is obtained from
$\Delta T$ induced by applied thermal current.
Since there is no ``constant thermal current source'',
the measured $\kappa$ always has some uncertainty coming
from the contact thermal resistance.
In contrast, we can use ``constant current source'' for $I_0$
in the Harman method, which is free from the contact thermal resistance
like a four-terminal method for resistance measurement.

Now we will evaluate the heat loss from current wires.
For simplicity, let us assume that the thermocouples are pasted 
just on the edges of the sample.
Then, in a steady-state, the thermal-current balance on the left edge 
is written as
\begin{equation}
 ST_LI_0=K\Delta T + I_0^2R/2 + K_0(T_0 - T_L) ,
\end{equation}
where $R$ is the resistance (including the contact resistance),
$T_0$ is the temperature at the heat sink,
and $K_0$ is the thermal conductance of the wire.
Here we take the current direction to give $T_R>T_L$ for $+I_0$.
Similarly, the thermal-current balance on the right edge is written as
\begin{equation}
ST_RI_0 = K\Delta T + K_0(T_R - T_0)-I_0^2R/2.
\end{equation}
Then we get
\begin{eqnarray}
I_0^2R + SI_0\Delta T &=& 2K_0(\bar T-T_0),\\
S\bar TI_0 &=& (K + K_0/2)\Delta T,
\end{eqnarray}
where $\bar T=(T_R+T_L)/2$.
These equations mean that $K$ and
$K_0$ are simultaneously determined 
by measuring $S$, $R$, $T_L$ and $T_R$.
In our setup, $K_0$ was estimated to be 2$\times$10$^{-5}$ W/K 
by calculating a thermal conductance of a gold wire 
with a 20-$\mu$m diameter and a 10-mm long at 300 K.
This was consistent with the values obtained through Eq. (4) from 
the measured values of $T_L$, $T_R$ and $T_0$,
and was subtracted at every measurement.
We evaluated  the radiation loss to be 
6$\times$10$^{-6}$ W/K at 300 K through the Stefan-Boltzmann law.
Since a typical value of $K$ is 3-5$\times$10$^{-4}$ W/K,
we can safely neglect the contribution of the radiation loss.

\begin{figure}[t]
\begin{center}
 \includegraphics[width=8cm,clip]{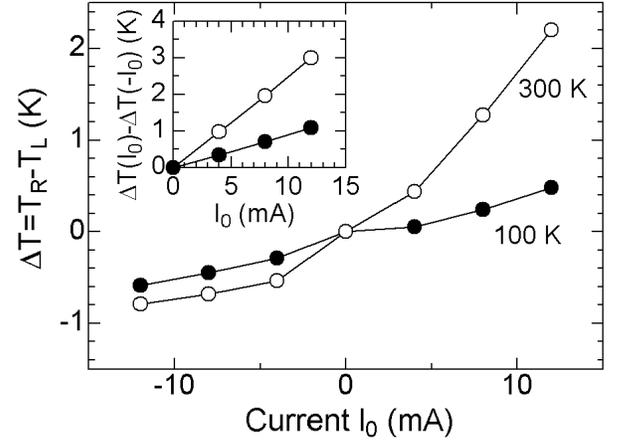}\\
\end{center}
\caption{
Temperature difference $\Delta T$ as a function of applied current $I_0$.
The inset shows anti-symmetric component of $\Delta T$.
}
\label{fig3}
\end{figure}

Figure 3 shows the observed $\Delta T$ as a function of $I_0$,
which contains a quadratic component of $I_0$ due to the Joule heating
from the misalignment of the thermocouples.
In order to cancel the quadratic term, we calculated
$\Delta T (I_0)-\Delta T(-I_0)$,
which is highly proportional to $I_0$,
as shown in the inset of Fig. 3.
We measured several samples to check the experimental accuracy, 
and found that the obtained $\kappa$ was reproducible within 
an error bar of less than 10 \% for
Bi$_{2-x}$Pb$_x$Sr$_2$Co$_2$O$_y$ and Ca$_3$Co$_4$O$_9$,
but was rather scattered for Na$_x$CoO$_2$.

Figure 4 shows the thermal conductivity of various layered
cobalt oxides along the CoO$_2$ block.
As shown in Fig. 4(a), the data for Na$_x$CoO$_2$ are
scattered from sample to sample.
This is partially because 
Na$_x$CoO$_2$ shows a smaller $S$ and weaker Peltier effect
than the other cobalt oxides. 
Another problem is the high contact resistance 
owing to the chemically active Na layer.
Nonetheless we can extract some common features from Fig. 4(a):
(i) $\kappa$ is highest among the three cobalt oxides.
(ii) $\kappa$ decreases with increasing temperature, indicating
a phonon-phonon scattering. 
(iii) The magnitude and temperature-dependence are
quite different from those for polycrystalline samples
\cite{takahata,takahata2}.
These features clearly indicate that the disordered Na layer 
is not very effective for the $\kappa$ reduction,
and that the grain boundary scattering 
seriously reduces $\kappa$ of polycrystalline samples.

Figure 4(b) shows $\kappa$ for Bi$_{2-x}$Pb$_x$Sr$_2$Co$_2$O$_y$
($x$=0, 0.4 and 0.6).
Unlike Na$_x$CoO$_2$, $\kappa$ decreases with decreasing temperature, 
and the magnitude is much smaller (particularly for $x=0$).
These results imply that the phonon mean free path 
is extremely short in Bi$_{2-x}$Pb$_x$Sr$_2$Co$_2$O$_y$.
Note that the solid solution between Bi and Pb increases 
$\kappa$, as is opposed to usual cases.
The Pb substitution decreases the excess oxygen \cite{karppinen},
and removes the superstructure in the Bi-O plane \cite{yamamoto}.
Thus the Bi-O plane becomes flatter for larger $x$, 
which would make phonon mean free path longer.
(The mass difference between Bi and Pb is so small that 
point-defect scattering due to Pb is negligible.)
Another notable feature is that a small but finite 
anisotropy exists between $a$ and $b$ axes.
We think that this is due to the misfit structure
between the hexagonal CoO$_2$ block
and the square Bi$_2$Sr$_2$O$_4$ block.
In fact, we found a large in-plane anisotropy in the resistivity and 
the thermopower of the same crystals \cite{fujii}. 

Figure 4(c) shows $\kappa$ for  Ca$_3$Co$_4$O$_9$.
As is similar to Bi$_{2-x}$Pb$_x$Sr$_2$Co$_2$O$_y$, 
a small in-plane anisotropy is observed,
which can be attributed to the misfit structure.
The magnitude and $T$ dependence are
intermediate between those for the previous two systems.
Note that a sharp drop below 50 K is 
clearly observed, which is a hallmark of ``clean'' crystals.
In this sense, the misfit structure does not
reduce $\kappa$ drastically, as was opposed to 
the theoretical prediction \cite{misfit}.

\begin{figure}[t]
\begin{center}
  \includegraphics[width=8cm,clip]{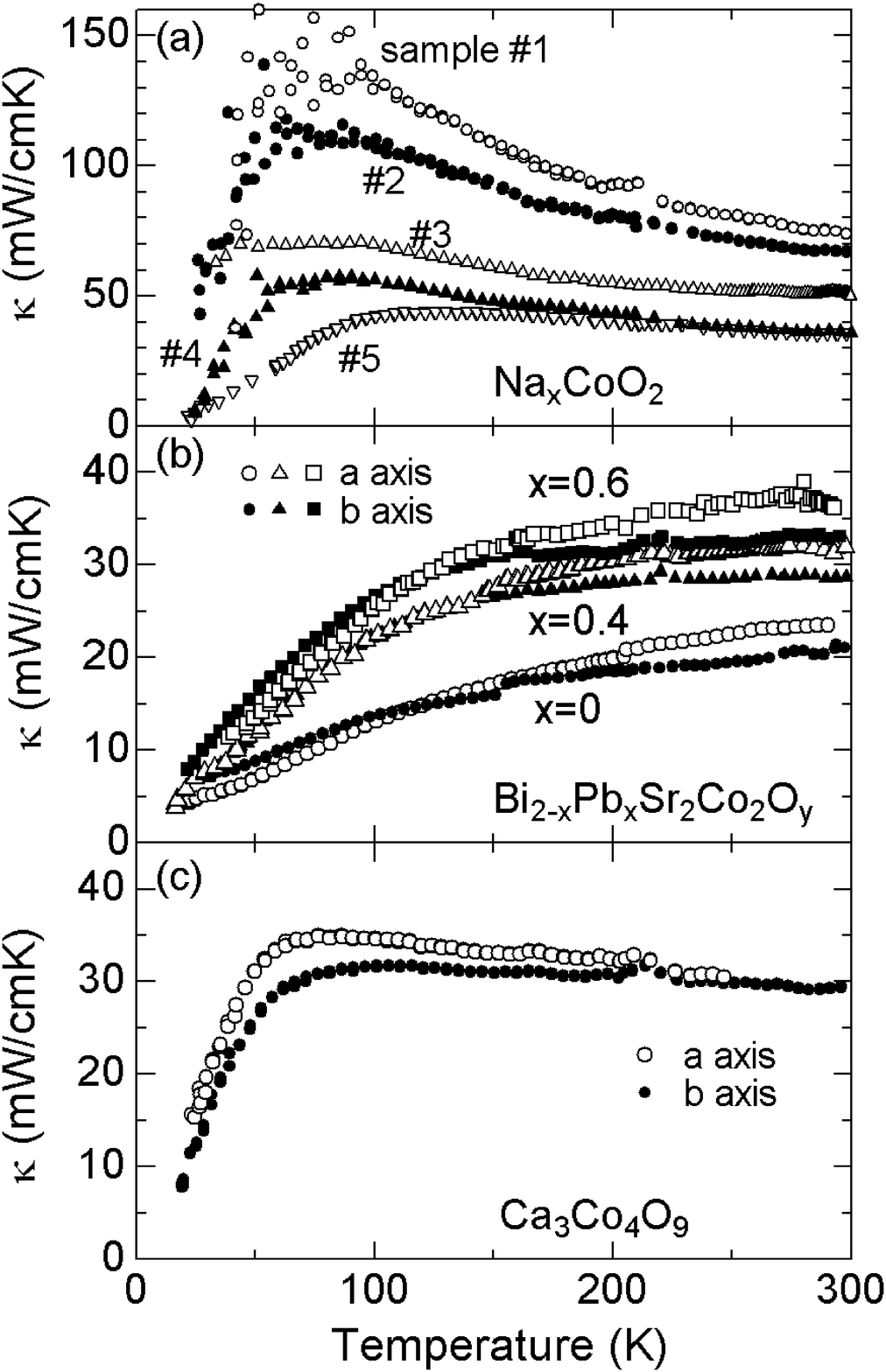}\\
\end{center} 
\caption{
The in-plane thermal conductivities of the layered cobalt oxides.
(a) Na$_x$CoO$_2$. Different marks represent different samples (\#1-\#5). 
(b) Bi$_{2-x}$Pb$_x$Sr$_2$Co$_2$O$_y$ ($x$=0, 0.4 and 0.6) and 
(c) Ca$_3$Co$_4$O$_9$. 
Open and filled symbols represent $a$- and $b$-axis thermal conductivity,
respectively.
}
\label{fig4}
\end{figure}

Let us compare the present data with the previous reports.
Funahashi and Shikano \cite{kappaBiCo,kappaCa} reported that 
the in-plane $\kappa$ of Bi$_{2}$Sr$_2$Co$_2$O$_y$ 
and Ca$_3$Co$_4$O$_9$ crystals
is 23 and 35 mW/cmK at 300 K, respectively,
which is in good agreement with our measurement.
This evidences the validity of the Harman method employed here.
For Na$_x$CoO$_2$, our data are qualitatively consistent 
with those reported by Fujita et al. \cite{fujita}, 
but the high $\kappa$ of Na$_x$CoO$_2$ seems overemphasized
in their measurement, presumably owing to the high
thermal resistance at the Na layer.
As mentioned above, the Harman method is free from contact (thermal) resistance,
and we believe that the present data are more correct. 
Very recently, Sales et al. have measured thermal conductivity 
of single-crystals of Na$_x$CoO$_2$, 
and reported nearly the same value as $\kappa$ of \#2 \cite{sales}.

Finally we will comment on the physical meaning
of the in-plane $\kappa$ in the layered cobalt oxides.
The present study has revealed the following features:
(a) Based on the Wiedemann-Franz law, the observed $\kappa$
is mainly due to phonon contribution.
(b) The lower $\kappa$ tends to show the weaker $T$-dependence.
(c) The lower $\kappa$ is seen in the material consisting of heavier atoms.
One may notice that Ca$_3$Co$_4$O$_9$ consisting of lighter atoms has a lower
$\kappa$ than Bi$_{1.4}$Pb$_{0.6}$Sr$_2$Co$_2$O$_y$.
This is perhaps because the Co-O layer in the Ca$_2$CoO$_3$ block is 
highly distorted owing to the misfit structure,\cite{miya2} 
and the acoustic phonon mode is less dispersive.

Considering these features, we can conclude 
that the in-plane $\kappa$ is mainly determined by
the phonons in the charge reservoir block 
(the block other than the CoO$_2$ block).
This implies that a nano-block integration concept works well
in the layered cobalt oxides, where {\it a nano-block
for the electronic part and a nano-block for 
the thermal part can be independently tuned or designed}.
This is a similar situation to the material design for high-temperature
superconductors, where a superconductor can be synthesized 
by combining proper charge reservoirs
with the superconducting CuO$_2$ plane \cite{arima}. 
As a result,
we can modify chemical properties, anisotropy and/or unit cell volume,
remaining the superconductivity intact.
Similarly, we can modify the thermal conductivity by changing the charge reservoir
without degrading the large thermopower in the CoO$_2$ block
in the layered cobalt oxides.

In summary, we have measured the in-plane thermal conductivity
of the thermoelectric layered cobalt oxides 
Na$_x$CoO$_2$, Bi$_{2-x}$Pb$_x$Sr$_2$Co$_2$O$_y$ and Ca$_3$Co$_4$O$_9$,
and have found that the charge reservoir block dominates the thermal conduction. 
By changing the charge reservoir layer, we can modify the thermal properties, 
remaining the electronic properties in the CoO$_2$ block unperturbed.

The authors would like to thank 
M. Shikano for helpful suggestion for crystal growth of Ca$_3$Co$_4$O$_9$. 
They also appreciate R. Kitawaki, W. Kobayashi
and K. Takahata for collaboration.

\end{document}